\documentclass[11pt]{revtex4-1}

\usepackage{graphicx}
\usepackage{dcolumn}
\usepackage{color}
\usepackage{bm}
\usepackage{amsmath}
\usepackage{amsfonts}
\usepackage{amssymb}

\input epsf

\begin{document}

\title{ Dynamical Systems analysis of an interacting dark energy model in the Brane Scenario}

\author{Sujay Kr. Biswas\footnote{sujaymathju@gmail.com}}
\affiliation{Department of Mathematics, Jadavpur University, Kolkata-700 032, India.}

\author{Subenoy Chakraborty\footnote{schakraborty@math.jdvu.ac.in}}

\affiliation{Department of Mathematics, Jadavpur University, Kolkata-700 032, India.}{}
\begin{abstract}
 In this paper, we investigate the background dynamics in brane cosmology  when dark energy is coupled to dark matter by a suitable interaction. Here we consider an homogeneous and isotropic Friedmann-Robertson-Walker ( FRW ) brane model and the evolution equations are reduced to an autonomous system by suitable transformation of variables. The nature of critical points are analyzed by evaluating the eigenvalues of linearized Jacobi matrix. Finally, the classical stability of the model is also studied.\\

Keywords: Braneworld cosmology; Dynamical system; Dark energy; Dark matter; Phase space.\\

PACS Numbers: 04.20.-q, 04.20.Cv, 98.80.-k ,98.80.Cq.
\end{abstract}
\maketitle
\section{Introduction}
In the brane world scenario [1,2] our universe is assumed to be a submanifold, or brane, embedded in a higher dimensional bulk space time. In contrast with the traditional Kaluza - Klein [3] treatment of extra - dimensions, ordinary matter fields are confined on the brane, only gravity can propagate into the bulk. In particular, the scenario proposed by Randall and Sundrum  [4], consists of  a single positive tension brane which is embedded in the five dimensional anti- de Sitter ( i.e. $\Lambda=-\frac{6}{\ell^{2}}$ , $ \ell $ is curvature radius of bulk ) space ( $ AdS_{5}$ ) with $ Z_{2}$  - symmetry ( the RS2 scenario  [4] ). However one may neglect the cosmological constant in $ 4D $ brane ( i.e. $ \Lambda_{4}=0 $ ) by choosing RS  fine tuning condition. It is well known that the cosmological field equations on the brane are essentially different from the standard 4 - dimensional cosmology [5].  In RS2  scenario, the modified Friedmann equations give  $ H\propto \rho_{T} $ at early stages of the evolution of the universe when the energy density is high ( i.e. $ \rho_{T} \gg \lambda $ ) and the standard four dimensional gravity is recovered on the brane in the low energy limit  [4,6,7] (i.e. $ H \propto\sqrt{\rho_{T}} $ when $ \rho_{T} \ll \lambda $ ).

Recent observations [8-14] suggest that our universe is currently undergoing an accelerating expansion. This challenging issue in standard cosmology shows a new imbalance in the governing Friedmann equations. Physicists have addressed such imbalances by either introducing new sources or by altering the governing equations. In the frame of standard cosmology, this imbalance is addressed by introducing a new matter source termed as dark energy (DE) in the Friedmann equations. However the nature of DE is completely unknown ( except for its negative pressure ) and is still an unresolved problem in modern cosmology [15,16]. On the other hand, a group of physicists have explored the second possibility and modified the gravity theory itself, assuming that at large scales, Einstein's theory of general relativity breaks down and a more general action describes the gravitational fields. These theories include f(R)- gravity, Scalar tensor gravity, Einstein-Gauss-Bonnet gravity, Brane world gravity and many others [2,4,15-21]. In particular, brane world scenario is related to gravity in higher dimension and matter fields are confined to the brane. In the frame work of RS2 model, several models have shown the recent observed phenomena, specifically a self-interacting scalar field [22-26] behaves as dark energy. The dynamics of scalar field with constant or exponential [27] as well as self-interaction potential [28] has been studied in the context of FRW cosmology. Also scalar field coupled to barotropic fluid has been found in [29-35]. Further, it should be mentioned that scalar field appears naturally in particle physics and in the present context it behaves as a source of DE.

 The aim of this paper is to investigate the dynamics of RS2 brane scenario in the context of interacting dark species. The dark energy (DE) is chosen as a real scalar field with self interacting potentials while a perfect fluid model with barotropic equation of state is taken as a model for dark matter (DM). The argument behind choice of interaction models is that they are favoured by observed data obtained from the Cosmic Microwave Background (CMB) [36] and matter distribution at large scales [37]. Further, Das et al [38] and Amendola et al [39] showed that an interaction model of the universe mimics the observationally measured phantom equation of state as compared to non-interacting models, which may predict a non phantom type of equation of state.

 As the evolution equations in the brane scenario are very complicated in form so exact analytical solution is not possible for the present model. Hence a dynamical system analysis in the phase space associated to this scenario is  presented. To study the nature of the critical points, eigenvalues of the first order perturbation matrix near the critical points are examined. Also classical stability analysis for the model is done. In this context, relevant dynamical system studies in the brane scenario can be found in [40-43] and others.

 The paper is organized as follows : Section II comprises of the essential details of Randall - Sundrum model and deals with basic equations in Brane Scenario and formation of Dynamical System. In Sections III a detailed phase space analysis related to the critical points is presented, we discuss the existence and stability / instability of critical points as well as of the model in section IV. Finally, the discussion and concluding remarks are given in section V. Throughout the paper we use natural units ( $ 8\pi G=\frac{8 \pi}{m_{PL}^{2}}=\hbar=c=1 $ ). \\
\section { Basic Equations in Brane scenario and formation of Dynamical System }

In the framework of RS2 brane model two dark species interacting non-gravitationally are considered in the background of flat FRW model. One of the dark species namely the dark matter is considered in the form of perfect fluid with barotropic equation of state while the other dark component namely the dark energy (DE) is chosen as real scalar field with arbitrary self interaction potential. So the modified Einstein field equations are  [44-47]
\begin{equation}
 H^{2} = { \frac{1}{3}}\rho_{T}(1+\frac{\rho_{T}}{2\lambda})+\frac{2U}{\lambda}
\end{equation}

\begin{equation}
  2 \dot{H} = - (1+\frac{\rho_{T}}{\lambda})(\dot{\phi}^{2}+\omega_{m}\rho_{m})-\frac{4U}{\lambda}
\end{equation}

Where $ \lambda $ is brane tension, $ \omega_{m} $ is barotropic index of the dark matter, $ \rho_{T}=\rho_{m} + \rho_{\phi} $ is the total energy density of the dark species. Here $ U(t) =\frac{C}{a(t)^{4}} $ is the dark radiation term which arises due to non- zero bulk Weyl tensor and C is constant parameter related to the black hole mass in the bulk. If the bulk is chosen as AdS- Schwarzschild then $ C\neq0 $  [45] while C vanishes  [47,48] for AdS bulk. For simplicity of calculation in the present work we shall choose the later bulk model so that $ C=0 $.

It should be noted that we have neglected the cosmological constant in the 4D brane ( i.e. $ \Lambda_{4}=0 $ ) by choosing Randall-Sundrum fine tuning condition. The energy conservation equations for individual dark sectors are given by
\begin{equation}
    \dot{\rho_{m}} +3H\omega_{m}\rho_{m}=Q
\end{equation}
and
\begin{equation}
    \dot{\rho_{\phi}} +3H(\rho_{\phi}+p_{\phi})=-Q
\end{equation}
where $ Q $ stands for the interaction between the two dark species ( i.e. dark matter and dark energy ). For the time being $ Q $ is unspecified only it is assumed that $ Q $ does not change sign during the cosmic evolution. In the above continuity equation (3) $ \rho_{m} $ is the energy density for dark matter and is related to the thermodynamic pressure by the relation $ p_{m} =(\omega_{m}-1)\rho_{m} $, where $ \omega_{m} $ is the barotropic state parameter bounded by the relation $ 0 \leq \omega_{m} \leq 2 $. However, $ \omega_{m} $ should be very close to unity, and even greater, nor less than 1, in order to satisfy the usual energy conditions imposed for dark matter. On the other hand, the energy density and pressure of the scalar field $ \phi $ are given by
\begin{equation}
    \rho_{\phi}=\frac{1}{2}\dot{\phi}^{2} + V(\phi) , p_{\phi} = \frac{1}{2}\dot{\phi}^{2} - V(\phi)
\end{equation}
with $ V(\phi) $ as the self interacting potential. Now using (5) in the continuity equation (4) the evolution of the scalar field is given by
\begin{equation}
    \ddot{\phi} + 3H\dot{\phi} + \frac{dV(\phi)}{d\phi}= -\frac{Q}{\dot{\phi}}
\end{equation}
At the early stages of the evolution of the universe  the energy density is high ( i.e. $ \rho_{T} \gg \lambda $ ) so from the modified Friedmann equation (1) $ H \propto \rho_{T} $. On the other hand, at late times, due to expansion the energy density of the matter falls off so that $ \rho_{T} \ll \lambda $ and we have $ H \propto \sqrt{\rho_{T}} $ i.e. the equation (1) behaves as the usual Friedmann equation at late stages of the evolution. Thus, although the brane effects are dominant at early epochs but the modified Friedmann equation (1) describes the evolution of the universe at all times.

Moreover due to complicated form of the evolution equations namely equations (1), (2), (3) and (6), it is not possible to have an analytic solution. Hence for a qualitative idea about the cosmological behavior we shall put the evolution equations into an autonomous dynamical system. For this, we introduce the new variables
\begin{equation}
    x=\frac{\dot{\phi}}{\sqrt{6}H}  ,  y= \frac{V}{3H^{2}}  ,  z= \frac{\rho_{T}^{2}}{6\lambda H^{2}}
\end{equation}
which are normalized over Hubble scale. As a result, the evolution equations reduce to the following autonomous system of ordinary differential equations ( after some algebra )
\begin{multline}
   \frac{dx}{dN}=\sqrt{\frac{3}{2}}ys - 3x + \frac{3}{2}x^{3} \frac{(1+z)}{(1-z)} (2-\omega_{m})+ \frac{3}{2}x\omega_{m}(1-y-z) \frac{(1+z)}{(1-z)} -\alpha   \frac{(1-x^{2}-y-z)}{2x} \\
   \frac{dy}{dN}=-\sqrt{6} xys + 3y\frac{(1+z)}{(1-z)}[x^{2}(2-\omega_{m}) + \omega_{m} (1-y-z)] \\
   \frac{dz}{dN} = -3z[x^{2}(2-\omega_{m}) +\omega_{m} (1-y-z)] \\
   \frac{ds}{dN} = -\sqrt{6} xs^{2}f(s)\\
 \end{multline}
where we have introduced another dynamical variable related to the self interacting potential of the scalar field as
\begin{equation}
    s=-\frac{1}{V} \frac{dV}{d\phi}
\end{equation}
with
\begin{equation}
    f(s)=V\frac{\frac{d^{2}V}{d\phi^{2}}}{(\frac{dV}{d\phi})^{2}} -1
\end{equation}
and the independent variable is chosen as $ N=\ln a $.\\

In deriving the above autonomous system of ordinary differential equation we choose the interaction to be in the form $ Q=\alpha H\rho_{m} $, with $ \alpha $ as the coupling parameter. Usually, in the literature $ \alpha $ is chosen to be positive. This indicates that there is energy flow from DE to DM, as required to alleviate the coincidence problem and to satisfy the second law of thermodynamics. Hence in view of coincidence problem, the positive coupling is very reassuring. On the other hand, for negative coupling parameter there is decay of DM into DE. Such models allow for the possibility that there is no DE field in the very early universe and that DE 'Condenses' as a result of the slow decay of DM [49]. Also recently, it has been shown [50] that the coupling parameter is weakly constrained to negative values by Planck measurements. However, the negative coupling can not be counted to resolve the tension between the Planck and HST measurements of the Hubble parameter[50]. Although, the negative coupling does not help to alleviate the coincidence problem, it appears in the observed data fittings that models with negative coupling show most significant departure from zero coupling.

Moreover, from the view point of curvature perturbation, it has been shown that[51] when the interaction is proportional to the DE energy density, we get a stable curvature perturbation ( except $ \omega_{\phi}=-1 $ ) while for the choice of the interaction proportional to DM energy density or total energy density of the dark sectors, the curvature perturbation can only be stable when $ \omega_{\phi}< -1 $ ( $ \omega _{\phi} $ is the equation of state parameter for DE ).\\

Further, using the normalized variables into the first modified Friedmann equation, we obtain the density parameter for the dark matter as
\begin{equation}
    \Omega_{m}=1-x^{2} -y-z
\end{equation}
Due to the energy condition $ 0\leq \Omega_{m} \leq 1 $, the normalized variables are not independent but are restricted by the relation
\begin{equation}
    0\leq x^{2} +y + z \leq 1
\end{equation}
Also the ratio of the total energy density to the brane tension is given by
\begin{equation}
    \frac{\rho_{T}}{\lambda} = \frac{2z}{(1-z)}
\end{equation}
From the above relation we see that the early super dense region $ ( \rho_{T}\gg \lambda ) $ i.e. neighborhood of initial singularity corresponds to $ z=1 $ while $ z=0 $ indicates late time cosmological solution when $ \rho_{T} \ll \lambda $ ( i.e. low energy regime ). But the explicit form of the autonomous system shows that $ z=1 $ is not allowed by the system i.e. our model is not appropriate to describe the dynamics near the initial singularity ( possibly quantum effects will be appropriate ). However, from  mathematical point of view  the neighborhood of this initial singularity may be reached in the limiting sense ( i.e. asymptotically ). Thus the phase space of the above autonomous system can be described as
\begin{equation}
    \Omega_{\rho s} = [\{ x,y,z \} \times \{ s \}]
\end{equation}
With $ 0\leq x^{2} + y +z \leq 1 $, $ -1 \leq x \leq 1 $ , $ 0\leq y \leq 1 $ , $ 0 \leq z \leq 1 $ and $ s\in\Re $ .
Further, the cosmological parameters related to the scalar field namely the equation of state parameter ( $ \omega_{\phi} $ ) and the density parameter $ \Omega_{\phi} $ can be expressed by the newly defined variables as
\begin{equation}
    \omega_{\phi} = \frac{p_{\phi}}{\rho_{\phi}} =\frac{x^{2}-y}{x^{2}+y} , \Omega_{\phi} = \frac{\rho_{\phi}}{3H^{2}} = x^{2} + y
\end{equation}
Also the deceleration parameter has the explicit form
\begin{equation}
    q= -1-\frac{\dot{H}}{H^{2}}= -1 + \frac{3}{2}(\frac{1+z}{1-z})[x^{2}(2-\omega_{m})+\omega_{m}(1-y-z)]
\end{equation}

\section{Critical points and Phase -Space Analysis :}

The system of equations (8) forming an autonomous dynamical system has the following eleven critical points, which can be classified as
:\\

I. \textbf{Critical points} : $ P_{1} , P_{2} = (\pm 1,0,0,0)$, $ P_{8} = (0,1,0,0) $ \\

II. \textbf{Curves of Critical Points} : $ P_{3} = (x_{c},0,0,0) $, $ P_{4} , P_{5} = (\pm 1,0,0,s_{c}) $,
   $$ P_{7} =( 0,1-z, z\in[0,1) , 0 ) $$

III.  \textbf{Classes of Critical points } : $ P_{6} = (x_{c},0,0,s_{c}) $ , $ P_{9}=(0,0,1,s) $ ,
 $$ P_{10}=(\frac{3\omega_{m}-\alpha}{\sqrt{6}s_{c}} , \frac{2\alpha s_{c}^{2}+(3\omega_{m}-\alpha)^{2}(2-\omega_{m})}{6s_{c}^{2}\omega_{m}} , 0 , s_{c} ) $$
 $$ and  $$
 $$  P_{11}= (\frac{s_{c}}{\sqrt{6}}, 1-\frac{s_{c}^{2}}{6}, 0, s_{c} )$$\\

where $ s_{c} $ is a solution of $ f(s)=0 $ and $ x_{c}^{2}= -\frac{\alpha}{3( 2-\omega_{m} )} $. Note that as  $ 0 \leq\omega_{m} \leq 2 $ so for the existence of the critical points $ P_{3} $  and  $ P_{6} $ we must have $ \omega_{m} < 2 $ and $ \alpha  < 0  $. In other words, the fluid should not be ultra relativistic stiff fluid and the energy exchange should be in the reverse way for the critical points $ P_{3} $ and $ P_{6} $ to exist.

These critical points and relevant physical parameters at those points are shown in table I.

\begin{table}
\caption{ Table shows the location of the critical points and the value of the relevant physical parameters of those points of the autonomous system (8).}
\begin{tabular}{|c|c|c|c|c|c|c|c|c|c|}
 \hline
   $ P_{i} $ & $ x $ & $ y $ & $ z $ & $ s $ & $ \omega_{m} $ & $ \omega_{\phi} $ & $ \Omega_{m} $ & $ \Omega_{\phi} $ & $ q $\\\hline
   $ P_{1} $ & 1  &  0 & 0 & 0 & - & 1 & 0 & 1 & 2 \\\hline
   $ P_{2} $ & -1 & 0 & 0 & 0 & - & 1 & 0 & 1 & 2 \\\hline
   $ P_{3} $ & $ x_{c} $ & 0 & 0 & 0 & unrestricted & 1 & $ 1-x_{c}^{2} $ & $ x_{c}^{2} $ & $ q_{c} $ \\\hline
   $ P_{4} $ & 1 & 0 & 0 & $ s_{c} $ & - & 1 & 0 & 1 & 2 \\\hline
   $ P_{5} $ & -1 & 0 & 0 & $ s_{c} $ & - & 1 & 0 & 1 & 2 \\\hline
   $ P_{6} $ & $ x_{c} $ & 0 & 0 & $ s_{c} $ & unrestricted & 1 & $ 1-x_{c}^{2} $ & $ x_{c}^{2} $ & $ q_{c} $ \\\hline
   $ P_{7} $ &  0  & $(1-z)$ & $ z\in[0,1)$ & 0 & - & -1 &  0 & $ 1-z $ & -1 \\\hline
   $ P_{8} $ &  0  & 1 & 0 & 0 & - & -1 &  0 &  1 & -1 \\\hline
   $ P_{9} $ &  0  & 0 & 1 & $ s $ & - & undefined &  0 &  0 & undefined \\\hline
   $ P_{10} $ &$ \frac{3\omega_{m}-\alpha}{\sqrt{6}s_{c}} $  & $\frac{2\alpha s_{c}^{2}+(3\omega_{m}-\alpha)^{2}(2-\omega_{m})}{6s_{c}^{2}\omega_{m}} $ & 0 & $ s_{c} $ & - &$ \frac{\omega_{m}(3\omega_{m}-\alpha)^{2}}{(3\omega_{m}-\alpha)^{2}+\alpha s_{c}^{2}}-1 $ &$ 1-\frac{\alpha}{3\omega_{m}}$\\&&&&&&& $-\frac{(3\omega_{m}-\alpha)^{2}}{3s_{c}^{2}\omega_{m}}$ &$\frac{\alpha}{3\omega_{m}}$\\&&&&&&&& $+\frac{(3\omega_{m}-\alpha)^{2}}{3s_{c}^{2}\omega_{m}}$  &$ -1+\frac{3\omega_{m}-\alpha}{2} $\\\hline
   $ P_{11} $ &$\frac{s_{c}}{\sqrt{6}}$  &$ 1-\frac{s_{c}^{2}}{6}$ & 0 & $ s_{c} $ & - & $\frac{s_{c}^{2}}{3}-1 $ &  0 &  1 &$ -1+\frac{s_{c}^{2}}{2}$ \\\hline
  \hline
\end{tabular}
\end{table}


\begin{figure}
\centering
\begin{minipage}{.6\columnwidth}
\centering
  \includegraphics[width=1.0\columnwidth]{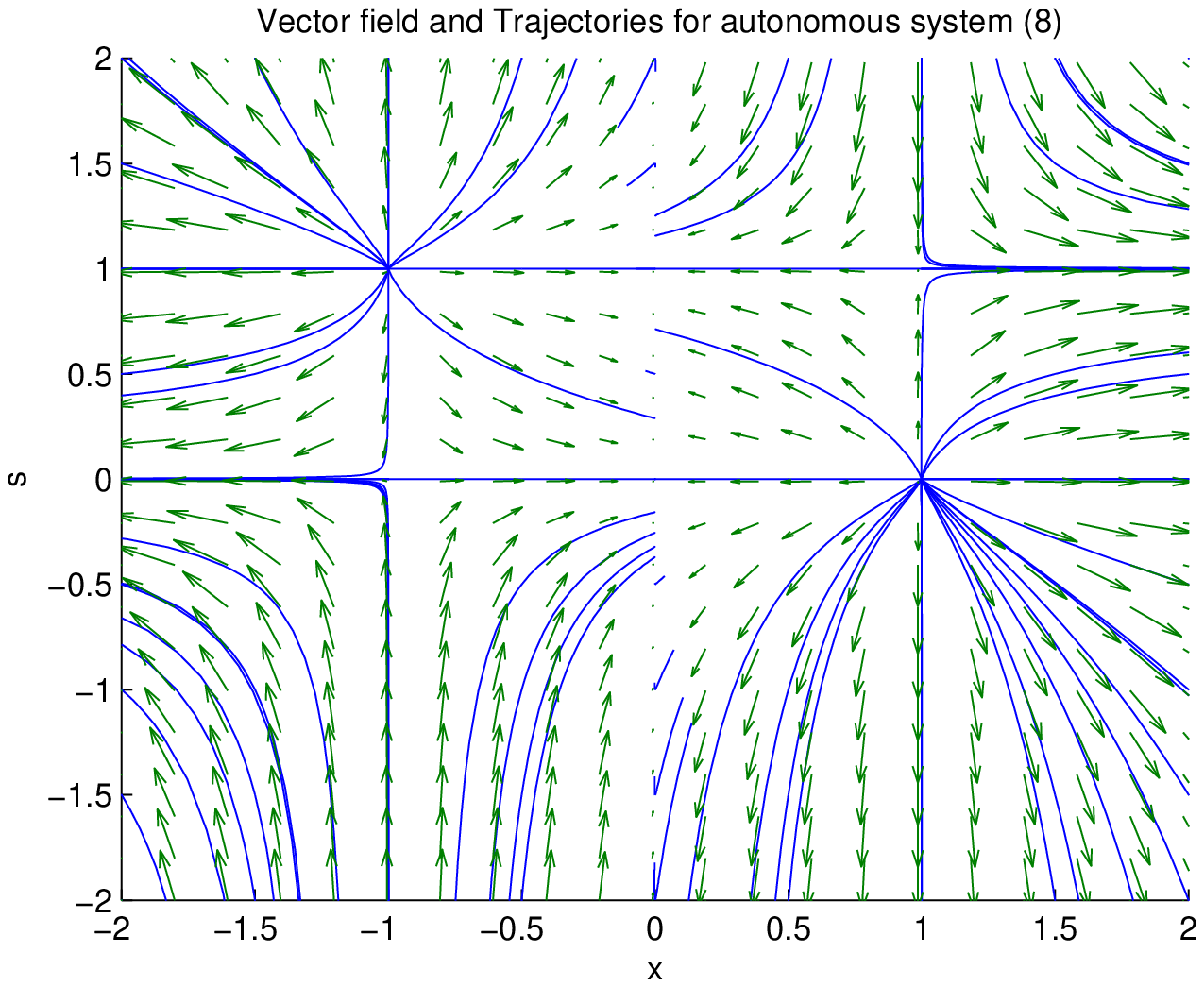}
 \caption{Phase space of the system (8)
 for the choices of
  ( $\alpha=0.01 $, $ \omega_{m}=1.01 $ , $\beta=1 $, $ \chi=-1 $ )
  for the self interaction potential $ V = \frac{V_{0}}{(\eta + \exp(-\chi\phi))^{\beta}}$ .}
  \label{fig:1}
\end{minipage}%
\hspace{0.7cm}
\begin{minipage}{.6\textwidth}
       \includegraphics[width= 1.0\linewidth]{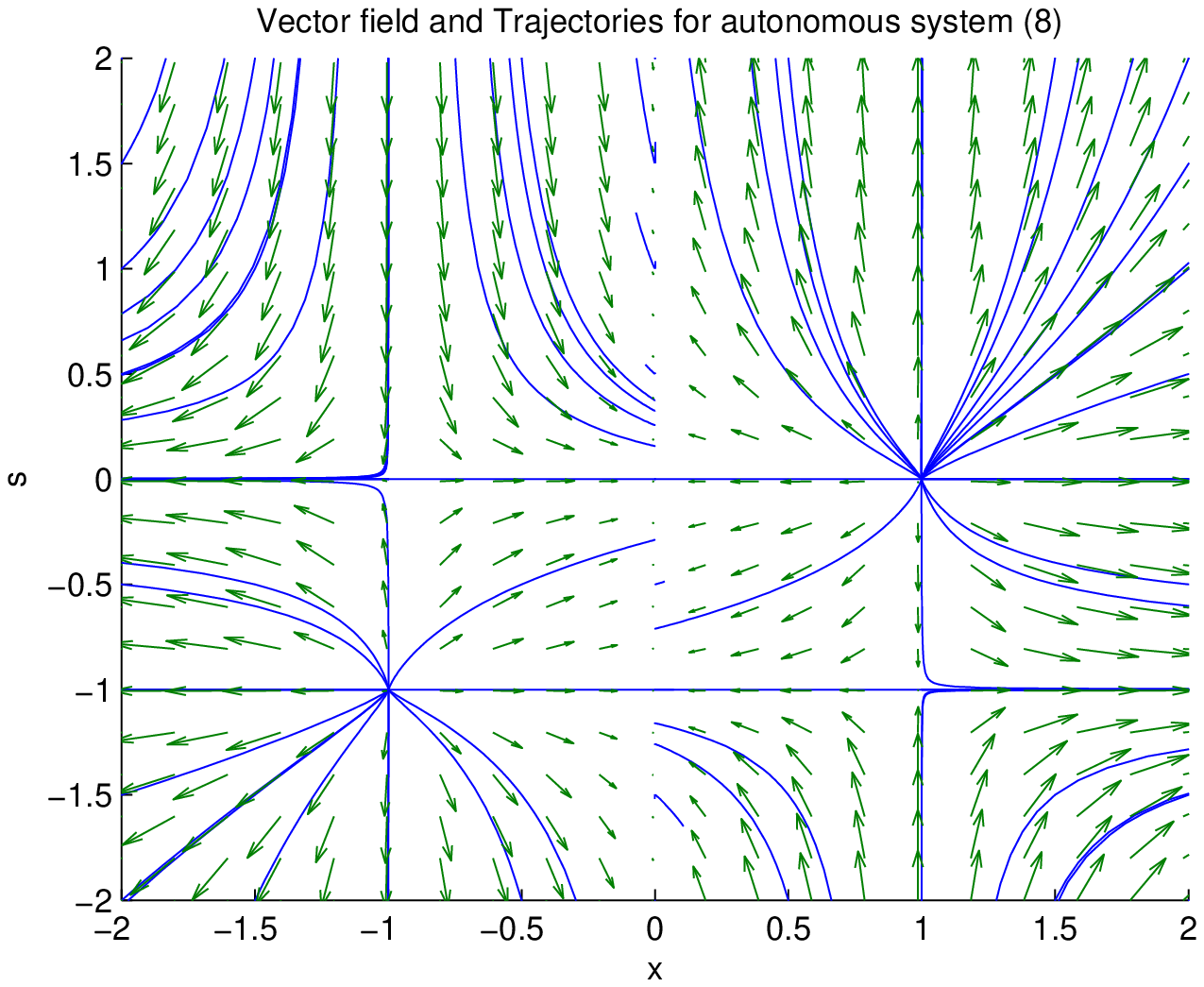}
       \centering

 \caption{Phase portrait  of the system (8) for the choices of
      $ \alpha=0.01 $ , $ \omega_{m}=1.01 $ , $ \beta=-1 $, $ \chi=-1 $ for the self interaction potential $ V = \frac{V_{0}}{(\eta + \exp(-\chi\phi))^{\beta}} $ }
\end{minipage}
\end{figure}
\begin{figure}
\centering
\begin{minipage}{.6\textwidth}
  \centering
  \includegraphics[width=1.0\linewidth]{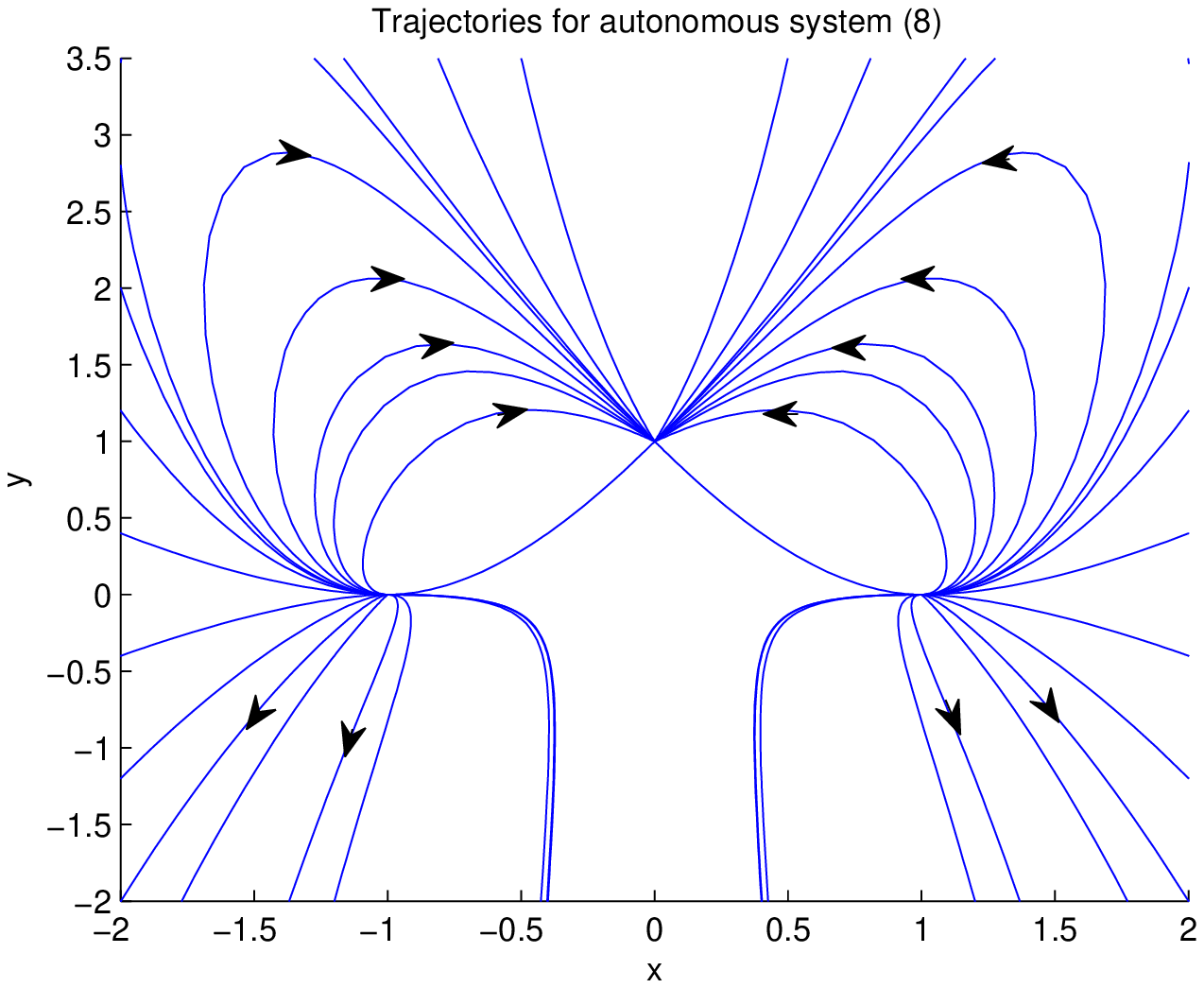}
  \caption{figure shows the the critical points $ P_{1} $ and $ P_{2} $ represent the unstable node in the (x, y) phase plane for $ \alpha=0.1 $ and $ \omega_{m}=1.01 $ while the point $ P_{8} $ represents the late time attractor inflationary de Sitter FRW solution
  .}
\end{minipage}%
\hspace{0.5cm}
\begin{minipage}{.6\textwidth}
  \includegraphics[width= 1.0\linewidth]{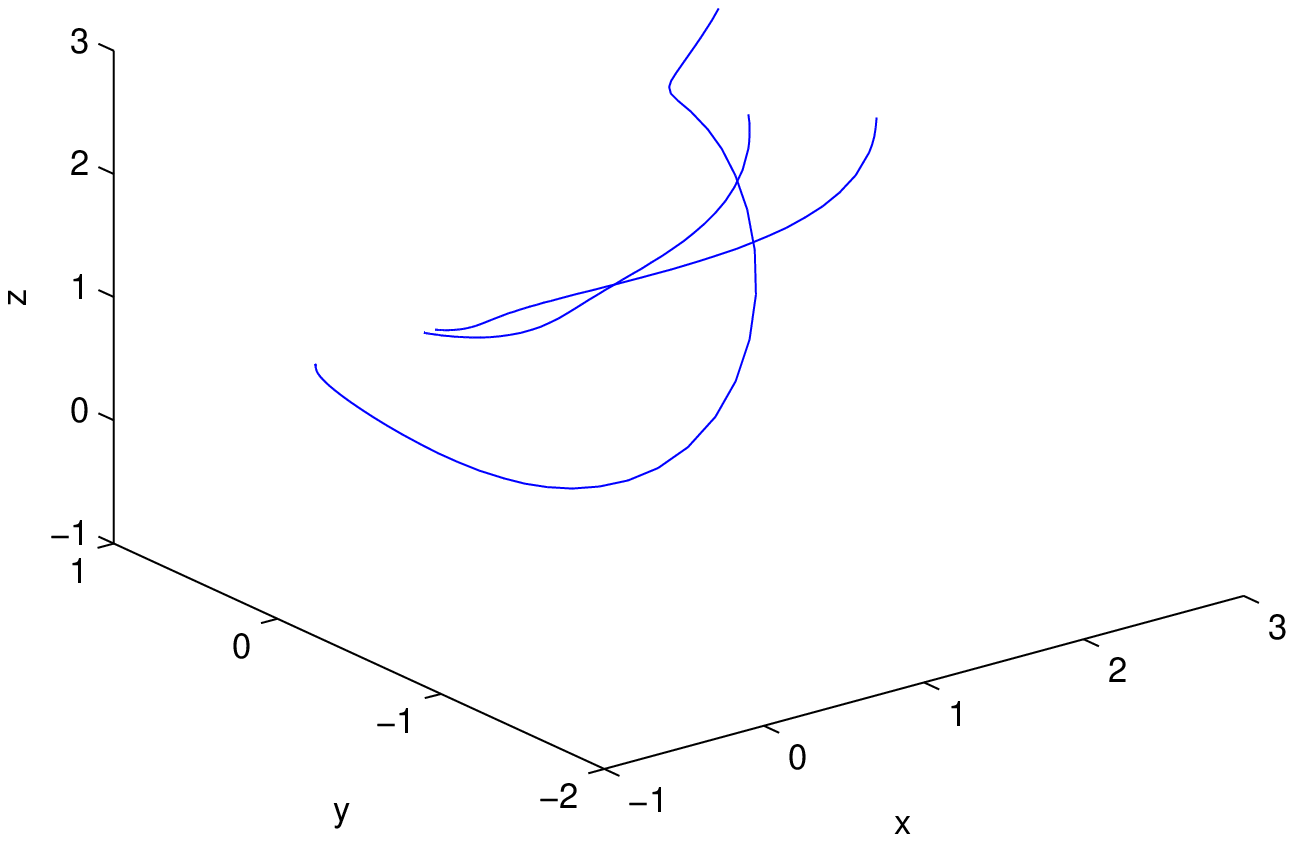}
\centering
  \caption{3D figure  of the system (8) for the choices of $\alpha=0.01 $ and $ \omega_{m}=1.03 $
    shows that the curve of the critical points $ P_{7} $ is a potential energy dominated attractor solution in the space. }
\end{minipage}
\end{figure}

\begin{figure}
\centering
\begin{minipage}{.6\textwidth}
\centering
\includegraphics[width=1.0\linewidth]{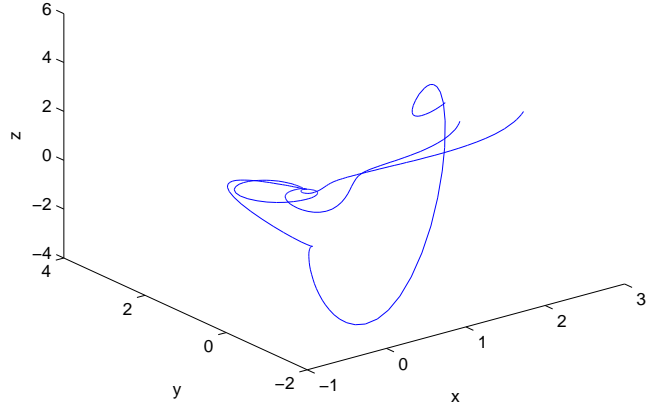}\\
\caption{ Figure of $ (x, y, z) $ of the system (8) for the choices of $ \alpha=0.1  $ and $ \omega_{m}=1.1 $  }
\end{minipage}
\end{figure}

\begin{figure}
\centering
\begin{minipage}{.6\textwidth}
\centering
\includegraphics[width=1.0\linewidth]{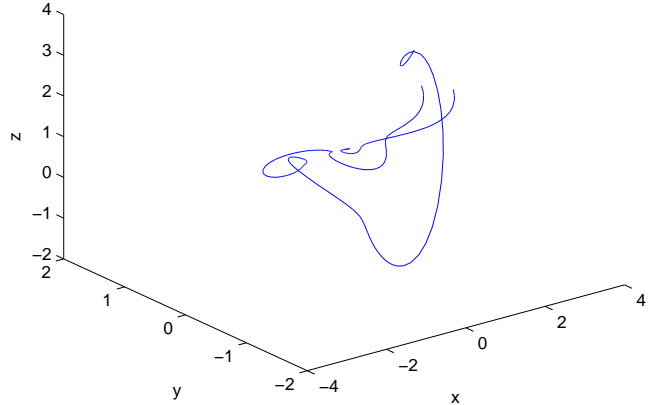}\\
\caption{ Figure of $ (x, y, z) $ of the autonomous system (17) for the choices of $ \alpha=1.1  $ and $ \omega_{m}=1.03 $  }
\end{minipage}
\end{figure}

\begin{figure}
\centering
\begin{minipage}{.6\textwidth}
\centering
\includegraphics[width=1.0\linewidth]{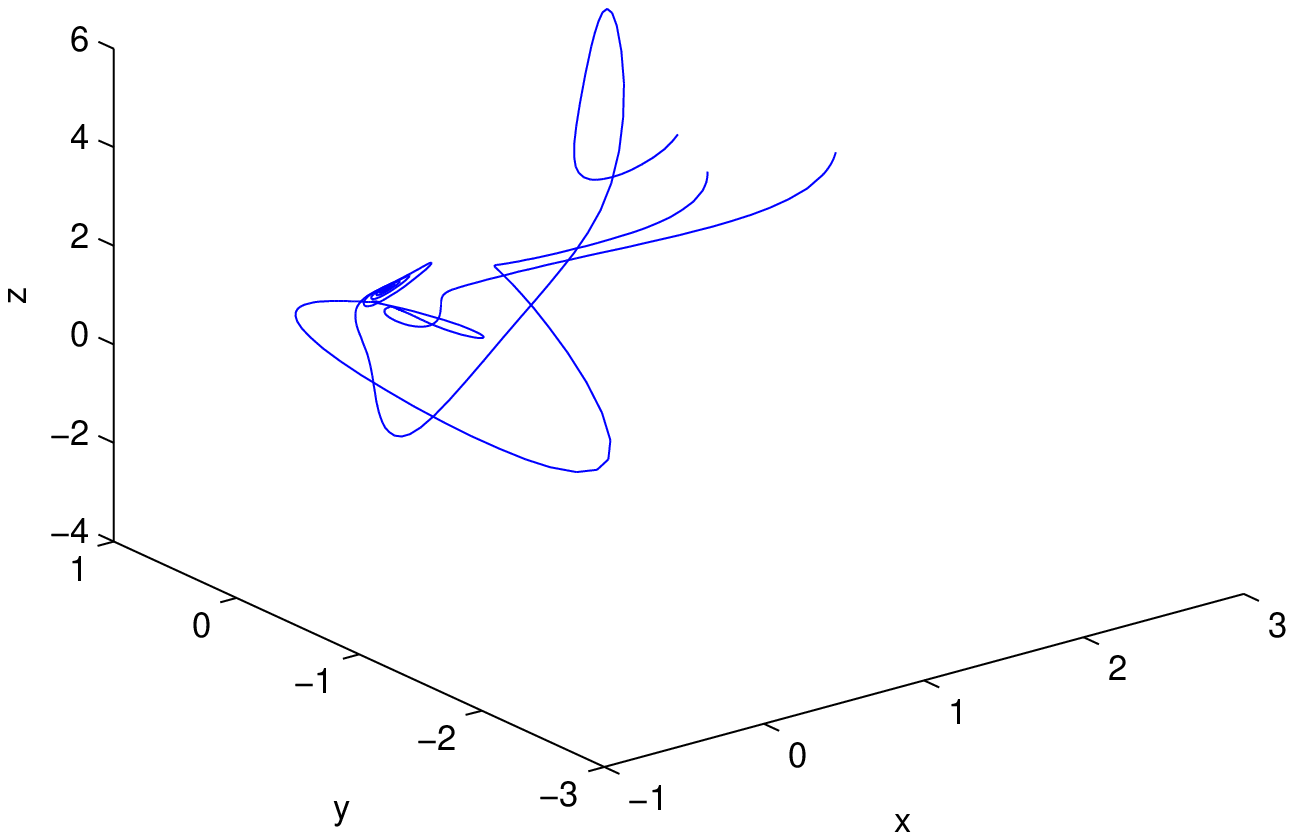}\\
\caption{ 3D figure of of the autonomous system (17) for the choices of $ \alpha=-1.5  $ and $ \omega_{m}=1.01 $  }
\end{minipage}
\end{figure}


 In the table $ q_{c} = -1 + \frac{1}{2} ( 3\omega_{m}- \alpha ) $.
 We see from the table that the critical points $ P_{1} $, $ P_{2} $, $ P_{4} $, $ P_{5} $, $ P_{7} $, $ P_{8} $ and $ P_{9} $ always exist while $ P_{3} $ and $ P_{6} $ will exist for $ x_{c}^{2}< 1 $  i.e. $ \omega_{m}<2 + \frac{\alpha}{3} $. Further the classes of critical points $ P_{10} $ will exist for $ s_{c}^{2}\geq 3\omega_{m}-\alpha $, $ 0\leq \omega_{m}\leq2 $ and for the coupling parameter $ \alpha> 0 $. These are the combination of both DE and DM and there will be an acceleration near $ P_{10} $ for $ 3\omega_{m}-\alpha<2 $. Another class of critical points $ P_{11} $ will exist for $ s_{c}^{2}\leq 6 $, completely DE dominated solutions and acceleration will occur near these points when $  s_{c}^{2}\leq 2 $. Also, the four critical points  $ P_{1} $, $ P_{2} $, $ P_{4} $  and $ P_{5} $  represent only the DE component ( DM is absent ) while $ P_{3} $ and $ P_{6} $ correspond to a combination of DM and DE with the ratio of two energy densities $ r= \frac{1-x_{c}^{2}}{x_{c}^{2}}$. Note that for the first six critical points ($ P_{1}$ - $ P_{6}$ ) the DE behaves as stiff fluid while for the critical points $ P_{7}$ and $ P_{8} $ the DE represents cosmological constant ($ \Lambda CDM $ model) and the DE corresponding to the critical points $ P_{10} $ and $ P_{11} $ is perfect fluid in nature. However for the critical point $ P_{9} $ we can not have any conclusion about the DE. For the two critical points $ P_{3} $ and $ P_{6} $ the perfect fluid representing the DM may have any equation of state. Although the four critical points $ P_{1} $, $ P_{2} $, $ P_{4} $  and $ P_{5} $ correspond to DE only but still in the brane scenario there will be deceleration only, while the critical points $ P_{3} $ and $ P_{6} $ represent a combination of DM and DE and there will be acceleration if $ \omega_{m} < ( \frac{2+\alpha}{3} ) $.
 Figures 1- 5 show the phase portrait ( in 2D and 3D ) of the system (8) for the self interaction potential  $ V = \frac{V_{0}}{( \eta + \exp(-\chi\phi ))^{\beta}} $  for different choices of the parameters involved. In this connection, it should be mentioned that a similar phase space analysis with exponential potential on the brane has been done by Goheer et al [52,53 ].\\

 \subsection{ Another Interaction model }

If we choose the Interaction term of the above 'interacting DE in the brane scenario' as $ Q= \alpha H \rho_{\phi} $ , where $ \alpha $ is coupling parameter.
The evolution equations reduce to the following autonomous system for this case :

\begin{multline}
   \frac{dx}{dN}=\sqrt{\frac{3}{2}}ys - 3x + \frac{3}{2}x^{3} \frac{(1+z)}{(1-z)} (2-\omega_{m})+ \frac{3}{2}x\omega_{m}(1-y-z) \frac{(1+z)}{(1-z)} -\alpha   \frac{(x^{2}+y)}{2x} \\
   \frac{dy}{dN}=-\sqrt{6} xys + 3y\frac{(1+z)}{(1-z)}[x^{2}(2-\omega_{m}) + \omega_{m} (1-y-z)] \\
   \frac{dz}{dN} = -3z[x^{2}(2-\omega_{m}) +\omega_{m} (1-y-z)] \\
   \frac{ds}{dN} = -\sqrt{6} xs^{2}f(s)\\
 \end{multline}
The critical points of the autonomous system (17) and the corresponding physical parameters are given in the table II:
\begin{table}
\caption{ Table shows the location of the critical points of autonomous system (17) and the values of the relevant physical parameters of those points.}
\begin{tabular}{|c|c|c|c|c|c|c|c|c|c|}
 \hline
   $ C_{i} $ & $ x $ & $ y $ & $ z $ & $ s $ & $ \omega_{m} $ & $ \omega_{\phi} $ & $ \Omega_{m} $ & $ \Omega_{\phi} $ & $ q $\\\hline
   $ C_{1} $ & $\sqrt{1+\frac{\alpha}{3(2-\omega_{m})}} $  &  0 & 0 & 0 & - & 1 & $-\frac{\alpha}{3(2-\omega_{m})}$ & $ 1+\frac{\alpha}{3(2-\omega_{m})} $ & $2+\frac{\alpha}{2} $\\\hline
   $ C_{2} $ & -$\sqrt{1+\frac{\alpha}{3(2-\omega_{m})}} $  & 0 & 0 & 0 & - & 1 & $-\frac{\alpha}{3(2-\omega_{m})}$ & $ 1+\frac{\alpha}{3(2-\omega_{m})} $ & $2+\frac{\alpha}{2} $ \\\hline
   $ C_{3} $ & 0 & 0 & 0 & 0 & - & undefined &  1 & 0 & $ -1+\frac{3\omega}{2} $ \\\hline
   $ C_{4} $ & 0 & 0 & 1 & 0 & - & undefined & 0 & 0 & undefined \\\hline
   $ C_{5} $ & 0 & 0 & 1 & $ s $ & - & undefined & 0 & 0 & undefined \\\hline
   $ C_{6} $ & $\sqrt{1+\frac{\alpha}{3(2-\omega_{m})}}  $ & 0 & 0 & $ s_{c} $ & - & 1 & $ -\frac{\alpha}{3(2-\omega_{m})} $ & $ 1+\frac{\alpha}{3(2-\omega_{m})} $ & $ 2+\frac{\alpha}{2} $ \\\hline
   $ C_{7} $ & -$\sqrt{1+\frac{\alpha}{3(2-\omega_{m})}} $   & 0 & 0 & $ s_{c} $ & - & 1 & $ -\frac{\alpha}{3(2-\omega_{m})} $  & $ 1+\frac{\alpha}{3(2-\omega_{m})} $ & $ 2+\frac{\alpha}{2} $  \\\hline
  \hline
\end{tabular}
\end{table}

From Table II, we see that the critical points $ C_{1} $ and $ C_{2} $ for the autonomous system (17) are same in all respect. They will exist for        $ \alpha<0 $ and $ 0\leq\omega_{m}\leq 2 $. They are combination of both DE and DM and are always dominated by the kinetic energy of the scalar field. There will be an accelerating phase of the universe near these critical points for $ \alpha<-4 $. On the other hand, the critical point $  C_{3} $ is completely matter dominated solution ($ \rho_{m}=3H^{2}$ ) ( see table II ) and are in accelerating phase for $ \omega_{m}<\frac{2}{3} $. Phase portrait of the autonomous system (17) are shown in figures 6 and 7 for different choices of the parameters involved.\\

\section{ Stability Analysis}

In this section, we shall investigate both the stability of the equilibrium points as well as the stability of the present model in two different subsections.

\subsection{ Equilibrium points and Stability criteria}

 We shall now discuss the stability of the critical points ( presented in table I ) of the autonomous system (8), considering first order perturbations near the critical points. To examine the nature of critical points, one has to study the eigenvalues of the first order perturbation matrix which has been presented in table III.  \\
 \begin{table}
\caption{ Eigenvalues of the linearized matrix for the critical points of the autonomous system (8).}
 \begin{tabular}{|c|c|c|c|c|}
   \hline
   $ P_{i} $ & $ \lambda_{1} $ & $ \lambda_{2}$ & $ \lambda_{3}$ & $ \lambda_{4}$ \\\hline

   $ P_{1} $ & $ 3(2-\omega_{m}) +\alpha $ & $ 6 $ & $ -6 $ &$  0 $ \\\hline

   $ P_{2} $ & $ 3(2-\omega_{m}) +\alpha $ & $ 6 $ & $ -6 $ & $ 0 $  \\\hline

   $ P_{3} $ & $ -\alpha-3(2-\omega_{m}) $ & $ -\alpha +3\omega_{m} $ & $ \alpha-3\omega_{m} $ & 0 \\\hline

   $ P_{4} $ & $ 3(2-\omega_{m}) +\alpha $ & $ -\sqrt{6}s_{c} +6 $ & -6 & $ -\sqrt{6}s_{c}^{2}f'(s_{c}) $ \\\hline

   $ P_{5} $ & $ 3(2-\omega_{m}) +\alpha $  & $ \sqrt{6}s_{c} + 6 $ & -6 & $ \sqrt{6}s_{c}^{2}f'(s_{c}) $ \\\hline

   $ P_{6} $ & $ -\alpha-3(2-\omega_{m}) $  & $ -\sqrt{6}x_{c}s_{c} + 2(1+q_{c})$  & $ -2(1+q_{c}) $  & $ -\sqrt{6}x_{c}s_{c}^{2}f'(s_{c}) $ \\\hline
   $ P_{7} $ & $ -3+\frac{\alpha}{2} $  & $ -3\omega_{m}(1+z) $  & $ 3 \omega_{m} z $  & 0 \\\hline
   $ P_{8} $ & $ -3+\frac{\alpha}{2} $  & $ -3\omega_{m} $  &  0   & 0 \\\hline
   $ P_{9} $ & $ 3(\omega_{m}-1)+\frac{\alpha}{2} $  & $ 6 \omega_{m} $  & $ 3 \omega_{m} $  & 0 \\\hline
   $ P_{10} $ & $ \frac{\alpha s_{c}^{2}}{\omega_{m}(3\omega_{m}-\alpha)}$\\&$-\frac{(3\omega_{m}-\alpha)(2-\omega_{m})}{2\omega_{m}} $  &$-\alpha$\\&&$-\frac{(2-\omega_{m})(3\omega_{m}-\alpha)^{2}}{2s_{c}^{2}} $ & $ -(3\omega_{m}-\alpha) $ & $-(3\omega_{m}-\alpha)s_{c}f'(s_{c})$ \\\hline
   $ P_{11} $ & $ -3+\alpha+\frac{s_{c}^{2}}{2}(3-\omega_{m}) $  & $\frac{\omega_{m}}{2}(s_{c}^{2}-6) $ & $ - s_{c}^{2}$  & $-s_{c}^{3}f'(s_{c})$ \\\hline
   \hline
 \end{tabular}
 \end{table}\\


The first two critical points $ P_{1} $, $ P_{2} $ are identical in all respects. Both are non-hyperbolic critical points but they behave like a saddle point in the phase space of the RS model because they have a non empty stable and unstable manifolds. These critical points represent solutions without matter part and are dominated by the kinetic energy of the usual scalar field.

 The line of critical point $ x=x_{c} $ ( $ P_{3} $ )  represents a solution which is a combination of both DM and DE.
 If $\mid x_{c}\mid $ is very close to unity then the solution is dominated by DE ( i.e. the scalar field ) while if $ x_{c} $ is very close to zero then it represents a DM dominated solution. It corresponds to a solution in accelerating or decelerating phase provided  $ \omega_{m}\lessgtr \frac{2+\alpha}{3} $. It is also non-hyperbolic in nature. However, it behaves like a saddle point in the phase space as it has a non empty stable and unstable manifolds.

 The critical point $ P_{4} $ is essentially a line of critical point $ s=s_{c} $ which are solutions dominated by the kinetic energy of the scalar field. These critical points are hyperbolic having saddle - like nature due to instability in the eigen direction associated with a positive eigenvalue and the stability of an eigen direction associated to a negative eigenvalue.

 $ P_{5} $ is another line of critical points characterized by $ s=s_{c} $ and has the same features as $ P_{4} $. Analyzing the eigenvalues we see that it is also hyperbolic saddle type critical points.

 Lastly, we have classes of critical points denoted by $ P_{6} $. Its corresponding solution is a mixture of DM and DE. However, depending on the choice of $ \omega_{m}  $  the solution may be in accelerated phase or in deceleration era. These hyperbolic critical points will be stable if $ \omega_{m} < min ( {\frac{\alpha}{3} + 2 , \frac{\alpha}{3} + \sqrt{\frac{2}{3}} x_{c}s_{c} } )$ and $ x_{c} f'(s_{c})>0 $, otherwise they will be saddle in nature. Note that in the stable case there may or may not be acceleration depending on $ x_{c}s_{c}\lessgtr \sqrt{\frac{2}{3}} $. However, if $ \omega_{m} > max ( {\frac{\alpha}{3} + 2 , \frac{\alpha}{3} + \sqrt{\frac{2}{3}} x_{c}s_{c} } ) $ and $ x_{c} f'( s_{c} ) < 0 $ then eigenvalues are positive and fixed points correspond to unstable solution.

 However, depending on the choice of parameters involved we have the following criteria
 $ x_{c} f'( s_{c} )>0 $ , $ -1 <q<\sqrt{\frac{3}{2}} x_{c}s_{c}-1 $ , the stable manifold is of dimension three and the system is restricted by the phantom barrier. Further, if the phantom region is allowed i.e. $ q <-1 $ then stable manifold is of dimension two. Moreover, if  $ s_{c} <0 $ and $ x_{c} f'( s_{c} ) <0 $ then all eigenvalues are positive and fixed points correspond to unstable solution.\\
\begin{table}
\caption{ Condition for stability at each equilibrium point of the autonomous system (8).}
\begin{tabular}{|c|c|c|c|c|c|c|c|c|c|}
 \hline
   $ P_{i} $ & x & y & z & s & local stability &  classical stability  \\\hline

   $ P_{1} $ & 1 & 0 & 0 & 0 & unstable(saddle) & stable  \\\hline
   $ P_{2} $ & -1 & 0 & 0 & 0 & unstable(saddle) & stable  \\\hline

   $ P_{3} $ & $ x_{c}$ & 0 & 0 & 0  & unstable(saddle) & stable if\\&&&&&& $ \alpha>\frac{6x_{c}^{2}}{x_{c}^{2}-1} $ \\\hline

   $ P_{4} $ & 1 & 0 & 0 & $ s_{c} $ & unstable(saddle) & stable  \\\hline
   $ P_{5} $ & -1 & 0 & 0 & $ s_{c} $ & unstable(saddle) & stable  \\\hline
   $ P_{6} $ & $ x_{c} $ & 0 & 0 & $ s_{c} $  & stable if \\ &&&&&$ \omega_{m}$ \\ &&&&& $< min ( {\frac{\alpha}{3} + 2, \frac{\alpha}{3} + \sqrt{\frac{2}{3}} x_{c}s_{c} } )$ \\ &&&&& and $ x_{c} f'(s_{c})>0 $  & stable if\\&&&&&& $ \alpha>\frac{6x_{c}^{2}}{x_{c}^{2}-1} $   \\\hline
   $ P_{7} $ & 0 & $ (1-z) $ & $ z\in[0,1)$ & 0  & unstable(saddle) & stable(limiting)  \\\hline
   $ P_{8} $ & 0 & 1 & 0 & 0 & attractor if $ \alpha<6 $ & stable(limiting)  \\\hline
   $ P_{9} $ & 0 & 0 & 1 & $ s $ & unstable(saddle) & stable(limiting)  \\\hline
   $ P_{10} $ & $ \frac{3\omega_{m}-\alpha}{\sqrt{6}s_{c}}$ & $ \frac{2\alpha s_{c}^{2}}{6s_{c}^{2}\omega_{m}}$ \\&& $ +\frac{(3\omega_{m}-\alpha)^{2}(2-\omega_{m})}{6s_{c}^{2}\omega_{m}}$ & 0 & $ s_{c}$ & stable if\\&&&&& $ 2\alpha s_{c}^{2} < (3\omega_{m}-\alpha)^{2}(2-\omega_{m}),$\\&&&&& $ 3\omega_{m}>\alpha $ \\&&&&& and $ s_{c}f'(s_{c})>0 $  & stable if\\&&&&&&$(3\omega_{m}-\alpha)^{2}(\omega_{m}-1)\geq\alpha s_{c}^{2}$ \\\hline
   $ P_{11} $ & $ \frac{s_{c}}{\sqrt{6}}$ & $ 1-\frac{s_{c}^{2}}{6}$ & 0 & $ s_{c} $  & stable if\\&&&&& $ \alpha< 3- \frac{s_{c}^{2}}{2}(3-\omega_{m})$\\&&&&& and $ s_{c}f'(s_{c})> 0  $ & stable if $ s_{c}^{2}\geq3 $ \\\hline
  \hline
\end{tabular}
\end{table}

The above results can be summarized as follows :\\

$\bullet$ For $ \alpha  > 0 $, the non-hyperbolic solutions $ P_{1} $ and $ P_{2} $ (in table I) are always unstable in the phase space and are in complete kinetic energy domination, are always decelerating $(q=2 ) $ ( see fig. 1, 2 and 3 ).

$\bullet$ For $ \alpha < 0 $ and $ \omega_{m}<2 $, the curves of critical points $ P_{3} $ are non-hyperbolic and have both combination of DE and DM and are always unstable (saddle) in the phase space. There will be an acceleration of the universe near the points $ P_{3} $ if $ \omega_{m}<\frac{2+\alpha}{3} $.

$\bullet$ For $ \alpha > 0 $, the hyperbolic solutions $ P_{4} $ and $ P_{5} $ (see table I) are unstable in the phase space. Though they are completely DE (kinetic energy of the scalar field ) dominated but only deceleration ($ q=2 $) possible. See fig. 1 for  $ \alpha =0.01 $ , $ \omega_{m}=1.01 $, $ \beta=1$, $ \chi=-1$ and potential $ V=\frac{V_{0}}{(\eta + exp(-\chi\phi))^{\beta}}$, the point $ P_{4} $ is saddle node where as $ P_{5} $ is unstable node in the phase space. On the other hand, for $ \alpha =0.01 $, $ \omega_{m}=1.01 $, $ \beta=-1$, $ \chi=-1 $ and potential $ V=\frac{V_{0}}{(\eta + exp(-\chi\phi))^{\beta}}$, the point $ P_{4} $ is saddle node and $ P_{5} $ is unstable node (see fig. 2 and tables I and III).

$\bullet$  For $ \alpha < 0 $ and $ \omega_{m}<2 $, the classes of critical points $ P_{6} $ are hyperbolic and combination of both DE and DM components. There will be an accelerated universe near $ P_{6} $ if $ \omega_{m}<\frac{2+\alpha}{3} $. These are the conditionally stable in the phase space.

$\bullet$ The solutions with 5D corrections in the brane scenario (singularities in the autonomous system (8)) namely $ P_{7} $, $ P_{8} $ and  $ P_{9} $ always exist. It should be mentioned that these singular points are similar to those in ref. [54 ] and the analysis is very similar to it. The de Sitter like solution ($ \omega_{\phi}=-1 $) $ P_{7} $ (see table I) is dominated by the potential energy of the scalar field ($ \rho_{T}=V $ and $\rho_{m}=0$ ) and is always in accelerating phase. Here potential much larger than brane tension i.e. $ V\gg \lambda $, so,  $ H_{RS}=\frac{V}{\sqrt{6\lambda}} $. So that the early time (high energy) expansion rate in the Randall-Sundrum model $ H_{RS} $ gets enhanced with respect to the general relativity rate $ H_{GR}$ :
$ \frac{H_{RS}}{H_{GR}}=\sqrt{\frac{V}{2\lambda}} $ and this is the kinetic energy dominated non-hyperbolic solution and has 2D stable manifold for $ \alpha <6 $, otherwise it is unstable node in the phase space ( see 3D fig. 4 and 5 ).

$\bullet$ The point $ P_{8} $ is particular case of $ P_{7} $ (in table I). $ P_{8} $ is non-hyperbolic in nature and has 2D stable subspace for $ \alpha <6 $. In particular, $ P_{8} $ corresponds to a late time attractor and de Sitter FRW solution ($ 3H^{2}=V $ ) (see figure 3).

$\bullet$  $ P_{9} $ is the kinetic energy dominated solution ($ \rho_{T}=\frac{\dot{\phi^{2}}}{2} $) which is also a saddle point for $ (3\omega +\frac{\alpha}{2}<3) $, non-hyperbolic in phase space. At first sight this is an unexpected result since in standard general relativity, the kinetic energy dominated solution is always a source ( past attractor ) in phase space. However, a closer inspection of the  RS model reveals that this point belongs in the high energy phase, which is the one that is modified by the brane contribution, hence it is not such an unexpected result; five dimensional contributions modify the structure of the phase space at high energies.

$\bullet$  For $ s_{c}^{2}\geq 3\omega_{m}-\alpha $, $ 0\leq\omega_{m}\leq2 $ and $ \alpha>0 $ the points $ P_{10} $ correspond to a combination of both DE and DM components and are hyperbolic in nature but may be non-hyperbolic for some choices of the parameters involved in the eigenvalues of corresponding critical points (see table III). The points will be stable in the phase space for $ 2\alpha s_{c}^{2} < (3\omega_{m}-\alpha)^{2}(2-\omega_{m})$, $ 3\omega_{m}>\alpha $ and $ s_{c}f'(s_{c})>0 $ otherwise they have stable manifold of dimension less than four (table III) . There will be an acceleration of the universe near $ P_{10}$ for $ 3\omega_{m}-\alpha<2 $.

$\bullet$ Finally, for $ s_{c}^{2}\leq 6 $, the classes of critical points $ P_{11} $  are completely DE ( scalar field, $ \Omega_{\phi}=1 $ ) dominated solutions (see table I) and will be stable in the phase space of the Brane Scenario for $ \alpha< 3- \frac{s_{c}^{2}}{2}(3-\omega_{m})$ and                      $ s_{c}f'(s_{c})> 0  $ (see table III). There exists an accelerating phase of the universe near $ P_{11} $ for $ s_{c}^{2}<2 $. $ P_{11} $ are hyperbolic in nature in the phase space but for some choices of $ s_{c} $ in table III they may behave like non-hyperbolic. Also it should be mentioned that the classes of critical points $ P_{10}$ and $ P_{11} $ are similar to critical points $ P_{7}$ and $ P_{8} $ in ref. [54].\\

 \subsection{ Stability of the model }

 In the present four dimensional autonomous system (8), the local stability criteria of an equilibrium point is characterized by the eigenvalues of the perturbation matrix ( presented in table III ) and discussion about local stability is presented in last subsection. We shall now investigate the classical stability of the model.

 In cosmological perturbation, sound speed ( $ C_{s} $ ) has a crucial role in characterizing classical stability. In fact, $ C_{s}^{2} $ appears as a coefficient of the term $ \frac{k^{2}}{a^{2}} $ ( $ k $ is the comoving momentum and '$ a $ 'is the usual scale factor ) and classical fluctuations may be considered to be stable when  $ C_{s}^{2} $ is positive[55,56] . This may prevent (to some extend) instability due to the presence of negative energy ghost states.
 In the present cosmological scenario we have

                     $$ C_{s}^{2} = 1 - \frac{2\sqrt{6}xys}{6x^{2}+\alpha(1-x^{2}-y-z)}  $$

 so for classical stability
\begin{equation}
            6x^{2}+\alpha(1-x^{2}-y-z) \geq 2\sqrt{6}xys
\end{equation}

  We have shown classical stability criteria of the model by the inequality (18). We shall now discuss about the criteria for the model stability at the equilibrium points ( presented in table I ) when x, y, z and s take the corresponding values of equilibrium points. From the tables I and III we see that the equilibrium points $ P_{1} $ and $ P_{2} $ are not locally stable ( saddle ) and from the above model stability analysis they correspond to classical stability. The equilibrium point $ P_{3} $ ( see table I ) corresponds to classical stability if $ \alpha > \frac{6x_{c}^{2}}{x_{c}^{2}-1} $  but it is not locally stable.
  From the tables I and III we see that two equilibrium points $ P_{4} $ and $ P_{5} $ are not locally stable ( saddle ) but from the above analysis we see that they are classical stable. The points $ P_{7} $, $ P_{8} $ and $ P_{9}$ correspond to classical stability (limiting). The classes of critical points $ P_{10} $ and $ P_{11} $ are classical stable conditionally (see table IV) as well as they are locally stable (see table III) which have been discussed earlier.
  Finally, the classes of critical points $ P_{6} $ locally stable if $ \omega_{m} < min ( {\frac{\alpha}{3} + 2 , \frac{\alpha}{3} + \sqrt{\frac{2}{3}} x_{c}s_{c} } )$ and $ x_{c} f'(s_{c})>0 $ which have been discussed earlier where as for classical stability they behave same as the equilibrium point $ P_{3} $. The corresponding condition for stability at each equilibrium point are presented in table IV.\\






\section{ Discussion and  Concluding Remarks }

 The present work deals with an explicit phase space analysis of the cosmological scenario in RS2 model. In the perspective of recent observational evidences, the matter is chosen in the form of interacting DE and DM. For DM, perfect fluid with barotropic equation of state is taken while real scalar field with self interacting potential is the candidate for DE. The relevant critical points of evolution equations ( which are hyperbolic or non-hyperbolic in nature ) and the values of the physical parameters at those points are presented in table I and table II for two type of interactions namely $ Q \propto \rho_{m} $ and $ Q \propto \rho_{\phi} $ respectively. The local stability of the equilibrium points are analyzed by examining the eigenvalues (presented in table III) of the linearized matrix. The classical stability of the system near the equilibrium points are discussed in section IV and the condition for classical stability is presented in the inequality (18). The self interacting potential for the scalar field (taken as DE) is chosen as $ V(\phi)=\frac{V_{0}}{(\eta+exp(-\chi\phi))^{\beta}} $.

 From the autonomous system (8) we have obtained two sets of critical points for different choices of the coupling parameter of the interaction term ( discussed in section III ). A wide class of self interaction potentials (DE) for which the quantity $ f(s) = \frac{VV''}{V'^{2}}-1 $ can be written as a function of variable $ s= -\frac{V'}{V} $ (where the primes denote differentiation with respect to $ \phi $ ) are included in this study.  For the self-interaction potential $ V(\phi)=\frac{V_{0}}{(\eta+exp(-\chi\phi))^{\beta}} $, the function $ f(s) $ is given by $ f(s) = \frac{1}{\beta} + \frac{\chi}{s} $ [57] and the zero of this function is $ s_{c} = -\beta\chi $. The phase portrait of system (8) for this potential are shown in figures 1 and 2.
 \\

 From the study we see that the critical points $ P_{1}$, $ P_{2} $, $ P_{3} $, $ P_{4} $ and $ P_{5} $  (hyperbolic or non-hyperbolic)  are saddle type  having union of nonempty stable and unstable manifolds. The critical points $ P_{1}$, $ P_{2} $, $ P_{4} $ and $ P_{5} $ are the  solutions dominated by kinetic energy of the scalar field but they correspond to decelerating phase ( $ q=2 $ ). There are some restrictions on $ \omega_{m} $ for accelerating or decelerating phase of the universe near the critical points $ P_{3} $ and $ P_{6} $. Also, we obtained the restrictions on $ \omega_{m} $ for which the classes of critical points $ P_{6} $ will be stable by analyzing the eigenvalues of the linearized perturbed matrix. The critical points $P_{7}$, $ P_{8}$ and $ P_{9} $ are non-hyperbolic in nature, among them $ P_{7} $ and  $ P_{8} $ have stable subspace of dimension 2 ( for $ \alpha<6 $ ) in the phase space and there  always exist accelerating phase of the universe near these points. The kinetic energy dominated solution $ P_{9}$ is unstable in the phase space. Further, classes of critical points $ P_{10} $ and  $ P_{11} $ are conditionally stable (see table III). Also in table II we have presented the critical points for the interaction $ Q \propto \rho_{\phi} $ and as the critical points are very similar so we have not discussed it further.\\

 Moreover, we have investigated the classical stability of the model. Note that these two type of stability are not interrelated because the stability of a critical point is related to the perturbations ( linear stability ) of corresponding point variables ( x, y, z and s ). On the other hand, the classical stability of the model is connected to the perturbations $ \delta p $ ( and depends on the condition $ C_{s}^{2}\geq 0 $ ). From the table IV, we conclude that the critical points can be classified into three categories namely \\
    (i) unstable points at which model is stable,\\
    (ii) stable points at which model is unstable and  \\
    (iii) stable points with stable model. \\

 The four critical points $ P_{1} $, $ P_{2} $, $ P_{4} $ and $ P_{5} $ are classical stable where as $ P_{7}$, $ P_{8} $ and $ P_{9} $ correspond to classically limiting stable and the equilibrium points $P_{10}$ and  $ P_{11}$ are both classically conditionally stable. The critical point $ P_{8} $ and  the classes of critical points $ P_{6} $, $ P_{10}$ and $ P_{11} $ are interesting from cosmological point of view. Imposing some restrictions on the independent parameters they will be stable points as well as correspond to stable model ( see table IV ) and describe late time acceleration for matter in the form of interacting dark energy.\\


 \begin{acknowledgements}

   One of the authors ( S C ) is thankful to UGC-DRS programme, in the Department of Mathematics, J.U. S C is also thankful to IUCAA, Pune for research facilities at Library.

 \end{acknowledgements}


%

\begin{thebibliography}{}

\bibitem{Akama1}

K Akama, Lect. Notes Phys. \textbf{176}, 267 (1982)\\
V A Rubakov and M E Shaposhnikov, Phys. Lett.\textbf{ B 125}, 139 (1983)\\
G W Gibbons and D L Wiltshire, Nucl. Phys.\textbf{ B 287}, 717 (1987)\\
P Horava and E Witten, Nucl. Phys. \textbf{B 460}, 506 (1996)\\
P Horava and E Witten, Nucl. Phys. \textbf{B 475}, 94 (1996)\\
N Arkani-Hamed, S Dimopoulos and G R Dvali, Phys. Lett. \textbf{B 429}, 263 (1998)\\
I Antoniadis, N Arkani-Hamed, S Dimopoulos and G R Dvali, Phys. Lett. \textbf{B 436}, 257  (1998)\\
N Kaloper, Phys. Rev.\textbf{ D 60}, 123506 (1999)\\
\bibitem{Randall1} L Randall and R Sundrum, Phys. Rev. Lett. \textbf{83}, 3370 (1999)
\bibitem{Ref}\label{3} T.Kaluza, Sitzungsber. Preuss. Akad. Wiss. Berlin (Math.Phys.)k1, 966(1921); O.Klein, Z.Phys. \textbf{37},895(1926).
\bibitem{Ref}\label{4} L Randall and R Sundrum, Phys. Rev. Lett. \textbf{83}, 4690 (1999)
\bibitem{Ref}\label{5} R.M.Hawkins , J.E.Lidsey, Phys. Rev. \textbf{D 63} , 041301 (2001)
\bibitem{Ref}\label{6} For a review on brane - world cosmology , see, e.g., D Langlois, Prog.Theor. Phys. Suppl. \textbf{\emph{148}}, 181(2003)
\bibitem{Ref}\label{7} J Garriga and T Tanaka , Phys. Rev. Lett. \textbf{84}, 2778(2000) \\
                       S B Giddings , E Katz , and L Randzll , J. High Energy , Phys.\textbf{ 0003}, 023 (2000)
\bibitem{Ref}\label{8} Adam G. Riess, Louis-Gregory Strolger, Stefano Casertano,
                       Henry C. Ferguson, Bahram Mobasher, et al. New
                       Hubble Space Telescope Discoveries of Type Ia Supernovae
                       at$ z >= 1 $: Narrowing Constraints on the Early
                       Behavior of Dark Energy. Astrophys.J., \textbf{659}:98-121, (2007).
\bibitem{Ref}\label{9} Tamara M. Davis, E. Mortsell, J. Sollerman, A.C. Becker,
                       S. Blondin, et al. Scrutinizing Exotic Cosmological Models
                       Using ESSENCE Supernova Data Combined with
                       Other Cosmological Probes. Astrophys.J., \textbf{666}:716-725, (2007) .
\bibitem{Ref}\label{10} W. MichaelWood-Vasey et al. Observational Constraints
                        on the Nature of the Dark Energy: First Cosmological
                        Results from the ESSENCE Supernova Survey. Astrophys. J., \textbf{666}:694-715, (2007).
\bibitem{Ref}\label{11} Max Tegmark et al. Cosmological parameters from SDSS
                        and WMAP. Phys.Rev., \textbf{D 69 }: 103501, (2004) .
\bibitem{Ref}\label{12} N. Jarosik, C.L. Bennett, J. Dunkley, B. Gold,
                        M.R. Greason, et al. Seven-Year Wilkinson Microwave
                        Anisotropy Probe (WMAP) Observations: Sky
                        Maps, Systematic Errors, and Basic Results. Astrophys. J.Suppl., \textbf{192} : 14 , (2011).
\bibitem{Ref}\label{13} D. Larson, J. Dunkley, G. Hinshaw, E. Komatsu, M.R. Nolta, et al. Seven-Year Wilkinson Microwave
                        Anisotropy Probe (WMAP) Observations: Power Spectra
                        andWMAP-Derived Parameters. Astrophys.J.Suppl., \textbf{192}: 16 , (2011).
\bibitem{Ref}\label{14} E. Komatsu et al. Seven-Year Wilkinson Microwave
                        Anisotropy Probe (WMAP) Observations: Cosmological
                        Interpretation. Astrophys.J.Suppl., \textbf{192} :18 , (2011) .
\bibitem{Ref}\label{15} T. P. Sotiriou and V. Faraoni : Rev. Mod. Phys. \textbf{82}, 451 (2010)
                        S. Nojiri and S.D. Odintsov : Phys. Rep. \textbf{505} , 59 (2011),
                        Ya-Bowu et al, Phys. Lett.  \textbf{B 717} 323 (2012)
\bibitem{Ref}\label{16} S. Capozziello : Int. J. Mod. Phys. Rev.   \textbf{D 11}   , 483 (2002)
\bibitem{Ref}\label{17} S. Nojiri and S.D. Odintsov : Phys. Rev.  \textbf{D 74}  , 086005 (2006)
\bibitem{Ref}\label{18} A. A. Starobinsky, Phys. Lett.   \textbf{B 91}   , 99 (1980) ,
                        R. Kerner : Gen. Relt. Grav. \textbf{ 14} , 453 (1982),
                        J. D. Barrow and A. Ottewi : J. Phys. \textbf{ A 16} , 2757 (1983)
                        V. Faraoni : phys. Rev. \textbf{ D 74}, 023520 (2006)
                        H. J. Schmidth , Int. J. Geom. Math. Phys.\textbf{ 4} , 209 (2007)
\bibitem{Ref}\label{19} S. Nojiri and S.D. Odintsov : Gen. Relt. Grav.  \textbf{ 36} , 1765 (2004),
                        Mod. Phys. Lett.  \textbf{ A 19} , 627 (2004);
                        M. C. B. Abdalla,  S. Nojiri and S.D. Odintsov : class. Quant. Grav. \textbf{ 22}, 235 (2005)
\bibitem{Ref}\label{20} S. Nojiri and S.D. Odintsov : Phys. Lett. \textbf{ B 576}  , 5 (2003);
                        S. M. Carroll et al., Phys. Rev. \textbf{ D 70}  , 043528 (2004) ;
                        S. Capozziello,  S. Nojiri and S.D. Odintsov : Phys. Lett. \textbf{ B 634} , 93 (2006)
                        S. Nojiri and S.D. Odintsov and D. Sacz-Gomes  : Phys. Lett. \textbf{ B 681} , 74 (2009)
\bibitem{Ref}\label{21} G. R. Bengochea and R. Ferraro : Phys. Rev. \textbf{ D 79}    , 124019 (2005)
\bibitem{Ref}\label{22} Kei-ichi Maeda. Brane quintessence. Phys.Rev.,\textbf{ D 64} : 123525, (2001).
\bibitem{Ref}\label{23} F Pedro . Gonzalez-Diaz. Quintessence in brane cosmology. Phys.Lett., \textbf{B 481} : 353 -359 , (2000).
\bibitem{Ref}\label{24} A.S. Majumdar.  Phys.Rev., \textbf{D 64} : 083503, (2001).
\bibitem{Ref}\label{25} N.J. Nunes and Edmund J. Copeland. Phys.Rev.,\textbf{ D 66 } : 043524, (2002).
\bibitem{Ref}\label{26} M. Sami and N. Dadhich. TSPU Vestnik, \textbf{44} N7 : 25-36, (2004).
\bibitem{Ref}\label{27} Tame Gonzalez, Tonatiuh Matos, Israel Quiros, and Alberto
                        Vazquez-Gonzalez. Phys.Lett., \textbf{B 676} :161-167, (2009).
\bibitem{Ref}\label{28} Yoelsy Leyva, Dania Gonzalez, Tame Gonzalez, Tonatiuh
                        Matos, and Israel Quiros. Phys.Rev., \textbf{D 80} : 044026, (2009).
\bibitem{Ref}\label{29} Edmund J. Copeland, Andrew R Liddle, and David
                        Wands. Phys.Rev., \textbf{D 57}: 4686-4690, (1998).
\bibitem{Ref}\label{30} J.M. Aguirregabiria and Ruth Lazkoz. Phys.Rev., \textbf{D 69} :123502, (2004).
\bibitem{Ref}\label{31} Ruth Lazkoz, Genly Leon, and Israel Quiros. Phys.Lett., \textbf{B 649}: 103-110, (2007).
\bibitem{Ref}\label{32} Wei Fang, Ying Li, Kai Zhang, and Hui-Qing Lu. Class.Quant.Grav., \textbf{26 }: 155005, (2009).
\bibitem{Ref}\label{33} Genly Leon. Class.Quant.Grav., \textbf{26}  :035008, (2009).
\bibitem{Ref}\label{34} Genly Leon, Pavel Silveira, and Carlos R. Fadragas.
                        Classical and Quantum Gravity: Nova Science Publishers,New York, USA,  arXiv:\textbf{ 1009.0689} [gr-qc] (2012).
\bibitem{Ref}\label{35} Genly Leon and Carlos R. Fadragas. Cosmological Dynamical
                        Systems. LAP Lambert Academic Publishing,
                        Germany, (2011).
\bibitem{Ref}\label{36} G. Oliveras, F. Atrio- Barandela and D. Pavon. Phys. Rev.\textbf{D 71} : 063523 (2005)
\bibitem{Ref}\label{37} G. Oliveras, F. Atrio- Barandela and D. Pavon. Phys. Rev.\textbf{D 74} : 043521 (2006)
\bibitem{Ref}\label{38} S. Das, P.S. Corasaniti and J. Khoury : Phys. Rev.\textbf{D 73} : 083501 (2006)
\bibitem{Ref}\label{39} L. Amendola, M.Gasperini and F. Plazza : Phys. Rev.\textbf{D 74} : 127302 (2006)
\bibitem{Ref}\label{40} A. Campos and C. F. Sopuerta : Phys. Rev.\textbf{D 63} : 104012 (2001)
\bibitem{Ref}\label{41} A. Campos and C. F. Sopuerta : Phys. Rev.\textbf{D 64} : 104011 (2001)
\bibitem{Ref}\label{42} A. A. Coley : Phys. Rev.\textbf{D 66} : 023512 (2002)
\bibitem{Ref}\label{43} A. A. Coley : astro-ph\textbf{/0504226}
\bibitem{Ref}\label{44} David Langlois. Brane cosmology: An Introduction.
                        Prog.Theor.Phys.Suppl., \textbf{148} : 181-212, (2003).
\bibitem{Ref}\label{45} Philippe Brax and Carsten van de Bruck. Class.Quant.Grav., \textbf{20} :R201- R232, (2003).
\bibitem{Ref}\label{46} David Langlois. Cosmology of brane - worlds.  arXiv: astro-ph/\textbf{0403579}  (2004).
\bibitem{Ref}\label{47} Roy Maartens. Brane world gravity. Living Rev.Rel., \textbf{7} :7, (2004).
\bibitem{Ref}\label{48} Peter Bowcock, Christos Charmousis, and Ruth Gregory. Class.Quant.Grav., \textbf{17} : 4745-4764, (2000).
\bibitem{Ref}\label{49} A. A. Costa , X. D. Xu, B. Wang, E. G. M. Ferreira and E. Abdalla : Phys. Rev.\textbf{D 89} : 103531 (2014)
\bibitem{Ref}\label{50} C. G. Boehmer, G. Caldera-Cabral, R. Lazkoz and R. Maartens :  Phys. Rev.\textbf{D 78} : 023505 (2008)
\bibitem{Ref}\label{51} Xi-ming Chen and Yungui, Emmanuel N. Saridakis \textbf{0812.1117} (arXiv) (2009)
\bibitem{Ref}\label{52} N. Goheer and P. K. S. Dunsby :  Phys. Rev.\textbf{D 67} : 103513 (2003)
\bibitem{Ref}\label{53} N. Goheer and P. K. S. Dunsby :  Phys. Rev.\textbf{D 66} : 043527 (2002)
\bibitem{Ref}\label{54} Dagobarto Escobar, Carlos R. Fadragas, Genly Leon, Yoelsy Leyva. arXiv: 1110.1736v3 [gr-qc].
\bibitem{Ref}\label{55} F.Piazza,  S.Tsujikawa ; JCAP \textbf{0407} , 004 (2004)
\bibitem{Ref}\label{56} N. Mahata, S. Chakraborty ; Gen. Relt. Grav. \textbf{46} , 1721 (2014) .
\bibitem{Ref}\label{57} Shuang- Yong Zhou. Phys. Lett., \textbf{B660} : 7-12, (2008).


\end{thebibliography}
\end{document}